# Title: Reduction of Alzheimer's disease beta-amyloid pathology in the absence of gut microbiota

**Authors:** T. Harach[1], N. Marungruang[2], N. Dutilleul[1], V. Cheatham[1], K. D. Mc Coy[3], J. J. Neher[4], M. Jucker[4], F. Fåk[2], T., Lasser[1] and T. Bolmont[1]

**Affiliations:**
[1] Laboratoire d'Optique biomédicale, Institute of Microengineering, School of Engineering, Ecole Polytechnique Fédérale de Lausanne, CH1015 Lausanne, Switzerland.
[2] Food for Health Science Centre, Lund University, Medicon Village, 22381 Lund, Sweden.
[3] Mucosal Immunology Lab, Department of Clinical Research, University of Bern, Murtenstrasse, 35 CH - 3010 Bern, Switzerland.
[4] German Centre for Neurodegenerative Diseases (DZNE), Tübingen, D-72076 Tübingen, Germany & Department of Cellular Neurology, Hertie Institute for Clinical Brain Research, University of Tübingen, D-72076 Tübingen.

## ABSTRACT

Alzheimer's disease is the most common form of dementia in the western world, however there is no cure available for this devastating neurodegenerative disorder. Despite clinical and experimental evidence implicating the intestinal microbiota in a number of brain disorders, its impact on Alzheimer's disease is not known. We generated a germ-free mouse model of Alzheimer's disease and discovered a drastic reduction of cerebral Aβ amyloid pathology when compared to control Alzheimer's disease animals with intestinal microbiota. Sequencing bacterial 16S rRNA from fecal samples revealed a remarkable shift in the gut microbiota of conventionally-raised Alzheimer's disease mice as compared to healthy, wild-type mice. Colonization of germ-free Alzheimer mice with harvested microbiota from conventionally-raised Alzheimer mice dramatically increased cerebral Aβ pathology. In contrast, colonization with microbiota from control wild-type mice was ineffective in increasing cerebral Aβ levels. Our results indicate a microbial involvement in the development of Alzheimer's disease pathology, and suggest that microbiota may contribute to the development of neurodegenerative diseases.

## INTRODUCTION

Alzheimer's disease (AD) is a severe and ever-growing socio-economic burden to western societies. According to the amyloid cascade hypothesis of AD pathogenesis, the aggregation and cerebral deposition of amyloid-β (Aβ) peptides into extracellular amyloid plaques is an early and critical event triggering a cascade of pathological incidents that finally lead to dementia [1]. Thus, arguing in favor of this hypothesis, the most rational strategy for an AD therapy would be to retard, halt and even reverse Aβ aggregation. However, despite all research efforts there is currently no treatment for AD, and all FDA-approved therapies only provide symptomatic treatments for this disease.

Numerous studies indicate that microbial communities represent an essential factor for many physiological processes including nutrition, inflammation, and protection against pathogens [2,3]. This microbial community is largely composed of bacteria that colonize all mucosal surfaces, with the highest bacterial densities found in the gastrointestinal tract. A growing body of clinical and experimental evidence suggests that gut microbiota may influence brain disorders including anxiety, autism and Parkinson's disease. The central nervous system's dysfunction in gamma-amino butyric acid (GABA)-mediated neurotransmission processes and brain-derived neurotrophic factor (BDNF) levels have both been connected to the development of anxiety. Gastro-intestinal bacteria are able to metabolize glutamate to produce the inhibitory neurotransmitter GABA [4]. The cerebral levels of the secreted protein BDNF, an essential component in neuronal survival, have also been correlated with alterations in profiles in gastro-intestinal microbiota [5,6]. Furthermore, recent pre-clinical research has revealed that microbiota may impact the development of autism spectrum disorders (ASD). *Bacteroides fragilis* improved defects in communicative and sensorimotor behaviors in the maternal immune activation in mice, a mouse model exhibiting ASD features [7]. Moreover, Scheperjans et al. highlighted a new connection between gut microbiota and Parkinson's disease [8], while Berer et al. demonstrated that in a mouse model of multiple sclerosis, gut microbiota strongly contributes to pathology[9]. All these findings strongly suggest that the gut microbiota may impact a wide range of brain functions. However, to date, the role of intestinal microbes on AD is largely overlooked and far from being understood.

## RESULTS

### Conventionally-raised APPPS1 animals display alterations in gut microbiota composition

Since microbial dysbiosis has been associated with many diseases we sought to determine whether conventionally-raised APPPS1 mice (CONVR-APPPS1)[10] developed changes in the intestinal microbiota as compared to healthy, wild-type age-matched littermate mice (CONVR-WT). Sequencing bacterial 16S rRNA genes extracted from fecal samples of 1, 3.5 and 8 month-old transgenic CONVR-APPPS1 and CONVR-WT mice revealed major shifts in the gut microbiota composition at both phylum and genus level in the 8 month-old transgenic CONVR-APPPS1 mice (**Fig. 1a, b**). Interestingly, younger CONVR-APPPS1 mice exhibited differences in gut microbiota as compared to age-matched CONVR-WT mice, although these alterations were less pronounced (**Fig. 1a, b**). In the 8 month-old mice, transgenic CONVR-APPPS1 mice displayed significant reductions in Firmicutes ($p < 0.001$), Verrucomicrobia ($p < 0.001$), Proteobacteria ($p < 0.01$) and Actinobacteria ($p < 0.01$) with a concurrent increase in the Bacteroidetes ($p < 0.001$) and Tenericutes ($p < 0.001$) phyla as compared to CONVR-WT mice. In the 1 month-old mice,

Bacteroidetes was increased in pre-depositing APPPS1 mice, while Cyanobacteria was decreased (**Fig. 1a, b**). In 3.5 month-old mice, similar changes in the gut microbiota were observed, but did not reach statistical significance (**Fig 1a**). Comparing high-abundant bacterial genera (relative abundance > 5 %) showed that the 8 month-old CONVR-APPPS1 had significantly increased abundance of unclassified genera in *Rikenellaceae* and *S24-7* (p< 0.001), while *Allobaculum* and *Akkermansia* (p< 0.001) were decreased as compared to CONVR-WT mice (**Fig. 1b**). An additional 10 low-abundant (< 5 %) genera differed significantly between CONVR-APPPS1 and CONVR-WT mice (data not shown). Furthermore, α-diversity was significantly higher in the 8 month-old CONVR-APPPS1 as compared to age-matched CONVR-WT mice (p=0.02) (**Fig. 1c**). Analysis of β-diversity using weighted (**Fig. 1c**) and unweighted (data not shown) UniFrac showed a clear, significant clustering of the aged mice depending on genotype (p<0.001). While body weight and food intake did not differ between groups, cecal weight was significantly higher in aged CONVR-APPPS1 mice as compared to age-matched CONVR-WT mice (p< 0.05) (**Fig. s3a, b**). We measured and observed a minor decrease in caloric content of the diet before and after autoclaving process (**Fig. s3b).**

**Generation of a novel Alzheimer transgenic model without gut microbiota**

Towards elucidating the role of the gut microbiota in AD, we generated a novel mouse model of AD without gut microbiota, referred to axenic, or germ-free (GF-) APPPS1 mice. The AD mouse model used for generation is derived from the APPPS1 transgenic mouse line [10]. APPPS1 animals co-express the KM670/671NL Swedish mutation of human amyloid precursor protein (APP) and the L166P mutation of human presenilin 1 (PS1) under control of the Thy-1 promoter, and show age-dependent accumulation of cerebral Aβ plaques. For germ-free rederivation, APPPS1 transgenic female mice were superovulated and bred with APPPS1 males. Plugged females were euthanized, oviducts collected, and the embryos were flushed out of the oviducts. The fertilized 2-cell embryos were collected and transferred into pseudo-pregnant germ-free NMRI recipient females under aseptic conditions. Resulting litters were maintained in flexible film isolators, and were regularly checked for their germ-free status. We compared GF-APPPS1 mice with conventionally-raised APPPS1 mice (CONVR-APPPS1) at different stages of amyloid pathology. In CONVR-APPPS1 mice, the first amyloid plaques appear in the cortex by 1.5 months and increase in number and size with ageing [10]. In the present study, we used young (3.5 month-old) and aged (8 month-old) amyloid-depositing GF-APPPS1 and CONVR-APPPS1 animals. Bacteriological evaluation of the GF-APPPS1 animals was regularly performed for their germ-free status by aerobic and anaerobic culture, DNA and gram staining of cecal content to detect uncultivable contamination. In addition, serological testing for known viruses and pathogens was performed periodically by ELISA, PCR or IFA tests (**Table s1 and s2**).

**GF-APPPS1 transgenic mice exhibit alterations of metabolic parameters**

Consistent with the absence of microbiota in GF-APPPS1 mice, cecum size was significantly smaller in young CONVR-APPPS1 mice (- 98.2 %, p< 0.001) (**Fig. s1a, b**). Furthermore, both young and aged GF-APPPS1 mice displayed significantly lower body weight compared to age-matched CONVR-APPPS1 counterparts (- 11.2 %, p< 0.01 and - 23.6 %, p< 0.001, respectively) (**Fig. s1c**). Accordingly, liver and stomach weights were significantly lower in young and aged GF-APPPS1 mice compared with their conventionally-raised counterparts (liver: - 33.3 %, p< 0.001 and - 14.2 %, p< 0.01, respectively; stomach: - 66.7 %, p< 0.001 and - 50.5 %, p< 0.01, respectively) (**Fig. s1d, e**). Interestingly, while pancreas weight was significantly lower in young

GF-APPPS1 mice (- 24.5 %, p< 0.05), it remained unaffected in aged GF-APPPS1 animals (p= 0.49) (**Fig. s1f**). Consistent with changes in body weight, fat pad weight was also significantly decreased in both young and aged GF-APPPS1 mice (- 57.5 %, p< 0.05 and - 58.3 %, p< 0.05, respectively) (**Fig. s1g**). Most surprisingly however, both young and aged GF-APPPS1 mice displayed significantly higher relative brain weight as compared to age-matched CONVR-APPPS1 counterparts (+ 33.3 %, p< 0.01 and + 27.3 %, p< 0.01, respectively) (**Fig. s1h**).

**Reduction of biochemical Aβ levels in GF-APPPS1 transgenic mice**

For neuropathological assessment, half a brain of each animal was prepared for biochemical evaluation (ELISA and western blot) and the other hemisphere processed for immuno-histological assessment. We first assessed levels of cerebral soluble human Aβ38, Aβ40 and Aβ42 by ELISA. In CONVR-APPPS1 mice, both Aβ40 and Aβ42 were detectable with Aβ42 exceeding Aβ40 several-fold, whereas Aβ38 was barely detected (**Fig. 2a-c**). All three Aβ species showed a significant level increase from 3.5 to 8 months of age (Aβ38: p< 0.001, Aβ40: p< 0.01, Aβ42: p< 0.01). In comparison to CONVR-APPPS1 mice, cerebral Aβ42 in GF-APPPS1 mice was significantly decreased in both young and aged animals (- 41.9 %, p< 0.01 and - 30.6 %, p< 0.01, respectively), whereas the decrease in Aβ40 levels did not reach statistical significance (p= 0.12 and p= 0.15, respectively). Aβ38 levels were unchanged (p= 0.14 and p= 0.32, respectively) (**Fig. 2b, c**). Cerebral Aβ42/Aβ40 ratio was significantly lower in young GF-APPPS1 mice in comparison to age-matched CONVR-APPPS1 animals (p< 0.01), but similar in aged APPPS1 animals (p= 0.53) (**Fig. 2d**). Consistent with the ELISA results, western blot assessment using mouse monoclonal anti-Aβ antibody 6E10 also demonstrated that soluble Aβ was significantly lower both in young and aged GF-APPPS1 mice compared to CONVR-APPPS1 animals (- 57.4 %, p< 0.001 and – 70.0 %, p< 0.001, respectively) (**Fig. 2e, f**). Interestingly, at least in the aged GF-APPPS1 mice, cerebral APP levels were significantly higher compared to CONVR-APPPS1 animals (+ 23.6 %, p< 0.05) (**Fig. 2e, f**). We then assessed plasma Aβ levels in GF- and CONVR-APPPS1 animals. While plasma levels of Aβ40 and Aβ42 were both significantly decreased in young GF-APPPS1 mice (- 20.0 %, p< 0.05 and - 29.4 %, p< 0.01, respectively) (**Fig. 2g, h**), plasma levels of Aβ40 and Aβ42 in aged animals remained unaffected (p= 0.66 and p= 0.75, respectively). Plasma Aβ42/Aβ40 ratio was significantly lower in young GF-APPPS1 mice in comparison to age-matched CONVR-APPPS1 animals (p< 0.01), but similar in aged APPPS1 animals (p= 0.28) (**Fig. 2i**).

**Decreased cerebral Aβ amyloidosis in GF-APPPS1 transgenic mice**

Histopathological assessment was performed to evaluate cerebral Aβ amyloid load in GF- and CONVR-APPPS1 animals. Brain sections from APPPS1 animals were stained with Thioflavin T dye (**Fig. 3a**). Quantification on Thioflavin T-stained brain sections revealed a significant reduction of compact Aβ amyloid load in both young and aged GF-APPPS1 brains (- 76.9 %, p< 0.001 and - 56.7 %, p< 0.001, respectively) as compared to age-matched CONVR-APPPS1 brains (**Fig. 3b**). Both young and aged GF-APPPS1 mice displayed significantly reduced compact pathology in the cortex (- 73.2 %, p< 0.01 and - 54.7 %, p< 0.001, respectively) and hippocampus (- 82.3 %, p< 0.05 and - 60.4 %, p< 0.01, respectively) (**Fig. 3b**). These results demonstrate that compact Aβ amyloidosis was significantly reduced in the absence of gut microbiota in GF-APPPS1 animals. Reduction of Aβ pathology in GF- versus CONVR-APPPS1 animals was confirmed qualitatively on transgenic brain sections immunostained with the anti-Aβ antibody 6E10. The reduction of cerebral Aβ amyloid load in GF-APPPS1 mice was accompanied by an overall decrease in cortical neuroinflammation, as evaluated by immunostaining with an antibody against the microglia marker Iba-1 (**Fig. s2a**). This observation was confirmed by quantitative imaging and revealed that both young and aged GF-APPPS1 animals exhibited a 64.2 % and 39.8 % decrease in Iba-1 positive immunostaining, respectively ($p < 0.001$ for both groups) as compared to age-matched CONVR-APPPS1 controls (**Fig. s2b**). To assess the neuroinflammatory profile, cytokines were measured in total brain homogenates from conventional and germ-free APPPS1 mice at 3.5 and 8 months of age (**Fig. s2c**). As expected, the pro-inflammatory cytokine interleukin (IL)-1β increased with age in CONV-APPPS1 mice by 117.2 % ($p < 0.001$). In contrast, IL-1β showed no significant increase with age in GF-APPPS1 mice ($p = 0.25$) and was 35.6 % lower in 8 month-old GF mice compared to CONV-APPPS1 animals ($p < 0.001$). Interestingly, three T cell-associated cytokines, IFN-γ, IL-2 and IL-5, were also significantly altered. IFN-γ, IL-2 and IL-5 levels were highest in 3.5 month-old CONV-APPPS1 mice and were significantly decreased by 46.1 % ($p < 0.01$), 43.6 % ($p < 0.01$) and 54.5 % ($p < 0.05$), respectively, in age-matched GF-APPPS1 mice. Immunostaining with only secondary anti-mouse antibody on GF-APPPS1 and CONVR-APPPS1 sections revealed no IgG leakage into the brain as evaluation of blood-brain-barrier leakage (data not shown).

**Upregulation of Aβ-degrading enzymes in GF-APPPS1 mice**

Several *in vitro* and *in vivo* studies have shown that Neprilysin degrading enzyme (NPE) and Insulin degrading enzyme (IDE) can degrade Aβ. As compared to age-matched CONVR-APPPS1 mice, levels of NPE were increased in both young and aged GF-APPPS1 animals by 116.1 % and 209.2 %, respectively ($p < 0.01$ and $p < 0.001$). Levels of IDE were increased in young GF-APPPS1 animals by 55.6 % ($p < 0.01$), but similar in aged GF-APPPS1 mice ($p = 0.43$), as revealed by western blot analysis using an anti-IDE antibody (**Fig. s4**). We also investigated levels of APP-CTF however no difference was found by western blot in young GF-APPPS1 animals as compared to age-matched CONVR-APPPS1 mice ($p = 0.08$). In contrast, levels of APP-CTF were increased in aged GF-APPPS1 animals ($p < 0.05$) (**Fig. s4**).

**Changes in the hypothalamic-pituitary-adrenal (HPA) axis in GF-APPPS1 animals**

GF-APPPS1 mice displayed several neuroendocrine changes in the hypothalamic-pituitary-adrenal (HPA) axis. Levels of Prolactin were significantly increased in aged GF-APPPS1 animals ($p < 0.05$) compared to age-matched CONVR-APPPS1 animals, but unaffected in young GF-APPPS1 mice ($p = 0.58$) (**Fig. s5a**). Despite a trend toward an increase, levels of thyroid stimulating hormone (TSH) remained unaffected in young and aged GF-APPPS1 mice ($p = 0.06$ and $p = 0.35$, respectively) compared to age-matched CONVR-APPPS1 animals (**Fig. s5b**). Levels of luteinizing hormone (LH) were increased in young and aged GF-APPPS1 mice ($p < 0.05$) but similar in aged GF-APPPS1 mice ($p = 0.96$) compared to age-matched CONVR-APPPS1 animals (**Fig. s5e**). Levels of Adrenocorticotropic hormone (ACTH) in young and aged GF-APPPS1 mice were unaffected compared to age-matched CONVR-APPPS1 animals ($p = 0.75$ and $p = 0.18$, respectively) (**Fig. s5d**). Levels of growth hormone (GH) in young and aged GF-APPPS1 mice were also unaffected ($p = 0.68$ and $p = 0.25$, respectively) compared to age-matched CONVR-APPPS1 animals (**Fig. s5e**).

**Colonization with microbiota from CONVR-APPPS1 mice but not from control wild-type mice increases Abeta pathology**

To further investigate the role of microbiota in the development of cerebral Aβ amyloidosis, 4 month-old GF-APPPS1 mice were colonized with the microbiota from aged CONVR-APPPS1 mice by oral gavage and were kept in a conventional environment (COLOAD-APPPS1 mice). Two months later, ELISA revealed that soluble Aβ38 and Aβ40 levels were unaffected between COLOAD-APPPS1 mice and control GF-APPPS1 mice ($p = 0.06$ and $p = 0.46$, respectively). In sharp contrast, levels of Aβ42 in COLOAD-APPPS1 mice were dramatically increased (+ 75.2 %) compared to control GF-APPPS1 mice ($p < 0.001$), without however reaching Aβ levels in CONVR-APPPS1 mice ($p < 0.05$) (**Fig. 4a-c**). Cerebral Aβ42/Aβ40 ratio was similar between COLOAD-APPPS1, GF-APPPS1, and CONVR-APPPS1 animals ($p = 0.09$) (**Fig. 4d**). Consistent with the ELISA results, western blot demonstrated that soluble Aβ was significantly higher in COLOAD-APPPS1 mice (+ 85.1 %, $p < 0.01$) as compared to GF-APPPS1 mice (**Fig. 4f**). Although approaching the level of Aβ in CONVR-APPPS1 mice, Aβ level in COLOAD-APPPS1 mice was 36.1 % lower ($p < 0.05$) (**Fig. 4e**). APP levels were significantly higher in GF-APPPS1 mice as compared to COLOAD-APPPS1 mice (+ 50.1 %, $p < 0.01$), but similar between COLOAD-APPPS1 mice and CONVR-APPPS1 mice ($p = 0.07$) (**Fig. 4e, f**). Plasma Aβ42, Aβ40 and Aβ42/Aβ40 ratios were similar between all mice groups ($p = 0.42$, $p = 0.07$ and $p = 0.38$, respectively) (**Fig. 4g, h, i**). In contrast and in comparison to microbiota from aged CONVR-APPPS1, control colonization of APPPS1 animals with microbiota from control wild-type mice (COLOWT-APPPS1) surprisingly lead to only few increase in Aβ levels / did not significantly increase Aβ levels.

**Comparison of gut microbiota composition in colonization experiments**

Bacterial 16S rRNA gene sequencing revealed that bacterial taxa in the Proteobacteria phylum initially colonized the mice (day 1-4), followed by an increase of taxa in Bacteroidetes, Firmicutes and Verrucomicrobia phyla (week 2-6) (**Fig. 5a**). There were no significant differences in gut microbiota composition between COLOAD-WT and COLOAD-APPPS1 mice at any of the different time points. Within both the COLOAD-WT and COLOAD-APPPS1 groups, there were significant differences in the microbiota composition over time (comparing day 1 with the other time points) (**Fig 5a**). Relative abundance of high-abundant (>5%) bacterial taxa at genus level in COLOAD-WT mice and in COLOAD-APPPS1 mice revealed that *Enterococcus*, *Parabacteroides* and unclassified genera in S24-7, *Enterobacteriaceae*, *Clostridiales*, *Lachnospiraceae* and *Rikenellaceae* differed over time ($p < 0.05$), as well as a range of low-abundant genera (**Fig. 5b**, **Fig. s6**). Alpha-diversity did not differ significantly between the COLOAD-WT mice as compared to the COLOAD-APPPS1 mice at any of the time points **(Fig. 5c)**. In contrast, β-diversity showed significant separation of COLOAD-APPPS1 and COLOAD-WT mice at day 1 ($p < 0.001$) (**Fig. 5c**), but not at later time points. An overview of correlations between the amount of cerebral soluble Aβ42 and the gut microbiota in aged APPPS1 and COLOAD-APPPS1 mice is shown in the OPLS plot (Fig.s7a). Importantly, the amount of cerebral soluble Aβ42 was positively correlated with the bacterial genera *Odoribacter* ($p < 0.001$), *Agrobacterium* ($p < 0.05$), *Anaerofustis* ($p < 0.05$), *Anaeroplasma* ($p < 0.05$), *Pseudomonas* ($p < 0.05$) and with unclassified genera in *Mogibacteriaceae* ($p < 0.05$), *Erysipelotrichiaceae* ($p < 0.05$), F16 ($p < 0.05$), *Xanthomonadaceae* ($p = 0.032$) and putative order *RF39* ($p < 0.001$) (**Fig. s7b**). Conversely, Aβ42 tended to be negatively correlated with *Parabacteroides* ($p = 0.067$) **(Fig. s7b).** Hence, both mice colonized with APPPS1 microbiota and aged CONVR-APPPS1 mice displayed associations between AD pathology and specific bacterial taxa.

## DISCUSSION

Increasing evidence suggests that gastro-intestinal microbiota is the living bridge between the gut and the central nervous system [11-13]. However the impact of gastro-intestinal microbes in the development of AD has, to the best of our knowledge, never been thoroughly investigated despite anecdotal evidence that amyloid plaque formation in APP transgenic mice differs among mouse facilities with different specific-pathogen-free (SPF) conditions. In order to elucidate the role of gut microbiota in the development of cerebral Aβ amyloidosis in AD we have generated an axenic mouse model of AD without gut microbiota (referred to as GF-APPPS1 mice). Herein, we showed that the absence of intestinal microbiota in the GF-APPPS1 transgenic model was sufficient to significantly decrease cerebral Aβ amyloid pathology. Both biochemical levels of Aβ and the extent of compact Aβ plaques were consistently decreased in the brains of APPPS1 transgenic mice without gut microbiota. Interestingly, the positive impact of the absence of microbiota on the progression of AD was already significant in young, 3.5 month-old GF-APPPS1 animals. The observation of similar or even higher APP levels in GF-APPPS1 mice compared to age-matched, CONVR-APPPS1 animals strongly implies that the decrease of Aβ amyloid pathology in the axenic AD animals is not due to lower APP expression, but rather results from a mechanism occurring downstream of APP production and cleavage. While our data on plasmatic Aβ42 levels in young GF-APPPS1 animals indicate that clearance of cerebral Aβ to the periphery is increased, the absence of such an effect in aged GF-APPPS1 animals suggests that increased efflux of Aβ from the brain does not significantly account for decreased Aβ amyloid pathology.

The marked differences observed in the gut microbiota composition between aged AD mice and healthy WT mice strongly advocate that a distinct microbial constitution in AD mice may play a role in the development of cerebral Aβ amyloidosis. Indeed, we demonstrated that the abundance of the two major phyla (Firmicutes and Bacteroidetes) in the fecal microbiota is dramatically altered by host genetics. At genus level, *Allobaculum* and *Akkermansia* were decreased in transgenic CONVR-APPPS1 mice and unclassified genera of *Rikenellaceae* and *S24-7* were increased. Reduced levels of *Akkermansia* have been associated with obesity and type 2 diabetes in mice [14], and the prebiotic-induced restoration of *Akkermansia* in the gut resulted in a lower fat-mass gain and decreased systemic inflammation, two risk factors in the development of AD pathology. Furthermore, the relative abundance of several bacterial genera was correlated with the amount of soluble Aß42 in the brain. Interestingly, this bacterial genus has previously been implicated in colon cancer and stress responses in mice [15,16]. The altered gut microbiota in the aged AD mice was not observed to the same extent in 1 month-old and 3.5 month-old transgenic CONVR-APPPS1 mice. However, the minor microbiota shift observed in the youngest, 1 month-old mice, gradually increased with age, indicating that bacteria might be involved in the progression of the disease. Interestingly, colonizing GF-APPPS1 mice with microbiota from CONVR-APPPS1 mice resulted in an intermediate microbial composition between CONVR-APPPS1 and WT mice, which correlated with intermediate levels of AD pathology in the COLOAD-APPPS1 mice as compared to CONVR-APPPS1 and GF-APPPS1 mice. Taken together, these results may support development of a novel microbial strategy for AD prevention and treatment.

To provide insights into the mechanisms by which Aβ amyloid pathology was decreased in GF-APPPS1 animals, we assessed the levels of Neprilysin degrading enzyme (NPE) and Insulin degrading enzyme (IDE), which normally degrade Aβ amyloid, resulting in lower cerebral Aβ accumulation. Most notably, NPE levels were increased in GF-APPPS1 mice of both ages, while IDE levels were increased in young GF-APPPS1 animals but similar in aged GF-APPPS1 mice as compared to age-matched CONVR-APPPS1 mice suggested that IDE contributes to decreased Aβ amyloid pathology at least in young mice. In contrast, and despite a strong trend toward an increase, similar levels of APP-CTF in young animals indicated that activity of gamma-secretase does not significantly account in differences observed in young GF- versus CONV-APPPS1 animals. However, the significant increase in APP-CTF in aged GF-APPPS1 suggest that gamma secretase activity might be altered in GF APPPS1 mice. Altogether, these results indicate that Aß degrading enzymes may partially play a role in decreasing Aß levels in young animals but is not sufficient to explain the overall drastic decrease of cerebral Aß in germ free animals.

Overall, GF-APPPS1 mice displayed various neuroendocrine changes in the hypothalamic-pituitary-adrenal (HPA) axis. Although levels of ACTH and GF were unaffected, aged GF-APPPS1 mice displayed increased Prolactin levels in aged GF-APPPS1 mice, suggesting that altered HPA axis might lead to stress and anxiety in these animals [17,18]. Increased levels of Prolactin are known to indirectly inhibit TSH and LH hormones. Consistently, TSH and LH remained unaffected in aged GF-APPPS1 animals. In contrast, levels of TSH and LH were increased in young GF-APPPS1 mice but comparable in aged GF-APPPS1 mice, revealing that the inhibitory effect of prolactin is alleviated in young GF-APPPS1 mice. These results indicate that HPA axis is altered in GF-APPPS1 mice and therefore suggest that anxiety phenotype might be enhanced in aged GF-APPPS1 mice. Changes in the HPA axis in young GF-APPPS1 mice were consistent with that previously observed in GF-WT type animals [19]. The results on neuroinflammatory profile are in line with a recent publication demonstrating that germ-free mice show immature microglia and reduced pro-inflammatory cytokine production [20]. Importantly, caspase-1 knockout (which prevents the production of IL-1β) has been shown to be sufficient to strongly reduce plaque load in APP/PS1 animals through altering the microglial activation state and enhancing microglial phagocytosis of Aβ plaques[21]. Therefore, a change in microglial responses in germ-free APPPS1 animals could contribute to the reduction of amyloid load observed in germ-free animals.

Our pre-clinical results using an axenic mouse model of AD demonstrate that the absence of microbiota retards substantially the progression of AD-like pathology. The gut colonization study underlines the important role of microbiota for the promotion of AD. The association of bacterial taxa with cerebral Aβ pathology observed in conventionally raised and colonized APPPS1 mice indicates that specific microbes may be involved in progression of AD pathology. Thus, our study strongly argues for a role of gastro-intestinal microbes in the development of cerebral Aβ amyloidosis. Obviously, the clinical translation of these preclinical results bears the potential for opening a new area for the treatment and prevention of AD pathology.

**Methods:**

*Mouse model of AD used for generation of axenic model:* APPPS1 transgenic mice were maintained at the Ecole Polytechnique Fédérale de Lausanne animal core facility. There was unlimited access to autoclaved food and water. APPPS1 animals co-express the KM670/671NL Swedish mutation of human amyloid precursor protein (APP) and the L166P mutation of human presenilin 1 (PS1) under the control of the Thy-1 promoter, and show age-dependent accumulation of parenchymal Aβ plaques with minimal vascular Aβ amyloid that is restricted to the pial vessels (*1*). APPPS1 mice were generated on a C57BL/6 background. Both male and female APPPS1 mice as well as age-matched control wild-type littermates were used. Both CONVR-WT and CONV-APPPS1 mice were housed together in grouped cages (n=5 mice/cage).

*APPPS1 germ-free rederivation:* For germ-free rederivation, 4-6 week-old APPPS1 transgenic female mice were superovulated by intraperitoneal injection of 5 IU pregnant mare serum gonadotropin (PMSG, VWR) on day 0, and 5 IU of human chorionic gonadotropin (hCG) on day 2. Females were paired with APPPS1 males on day 2. Two days later, plugged females were euthanized, oviducts collected, and the embryos were flushed out of the oviducts. The fertilized 2-cell embryos were collected and extensively washed in M2 medium (Sigma) containing Penicillin/Streptomycin (Invitrogen). Subsequently these embryos were transferred into pseudo-pregnant germ-free NMRI recipient females under aseptic conditions. Resulting litters were maintained in flexible film isolators at the Clean Mouse Facility at the University of Bern with unlimited access to autoclaved food and water. Both male and female GF-APPPS1 mice as well as age-matched control wild-type littermates (GF-WT) were used, and both CONVR-APPPS1 and GF-APPPS1 mice were fed with the same autoclaved diet (Kliba-Nafag PN3437). Both CONVR-APPPS1 and GF-APPPS1 mice were housed in grouped cages (n=5 mice/cage), and both CONVR-APPPS1 and GF-APPPS1 mice were handled 2 times a week.

*Bacteriological status of the GF-APPPS1 model:* Mice were regularly checked for germ-free status by aerobic and anaerobic culture, DNA staining (Sytox-green, Invitrogen) and gram staining of cecal content to detect uncultivable contamination. In addition, serological testing for known viruses and pathogens was performed periodically by ELISA, PCR or IFA. The obtained negative results confirm the germ free status of our animals (**Table s1 and s2**).

*Preparation of transgenic samples for analyses:* Animal studies were performed according to regulations issued by the Swiss government and approved by the ethic veterinary committee of the cantons of Vaud and Bern, Switzerland. Germ free and conventionally-raised animals were housed and bred according to standardized procedures. Germ free animals were maintained in flexible film isolators at the Clean Mouse Facility at the University of Bern with unlimited access to autoclaved food and water. One day prior to sacrifice, mice were shipped to EPFL in sterile containers. On the day of sacrifice, mice were weighed and euthanized by lethal intraperitoneal injection of pentobarbital (150 mg/kg; < 200 mg/ml); stomachs, livers, pancreas and brains were collected, weighed and snap frozen in liquid nitrogen for further analysis. Brains were extracted and cut in half; one half was immediately incubated in 4 % paraformaldehyde in ice-cold PBS and the other hemisphere was snap frozen until analysis.

*Histology & immunohistochemistry:* After 2 days fixation in 4% paraformaldehyde (PFA), the half brains were incubated for 48 h in 30% sucrose. Brains were then frozen in 2-propanol (Merck) and subsequently sectioned on a freezing-sliding microtome to collect 25 µm coronal sections. For histological staining, brain sections were stained with Thioflavin T dye according to standard protocol (ThT 1% in 50% Ethanol; Sigma Aldrich, St. Louis, MO). Sections were immunostained to visualize Aβ deposits using mouse monoclonal antibody (6E10, 1:500; COVANCE). Microglial reaction was assessed using a rabbit polyclonal antibody to Iba-1 (1:1000; Wako, 019-19141). Vectastain Elite ABC Kits were used and revelation obtained with Vector SG Blue (Vector Laboratories).

*Western blot & biochemistry:* Brains were extracted and homogenized in PBS (10% weight/volume). Brain homogenates were centrifuged at 14'000 rpm for 15 min at 4° C before the protein concentration was measured. The purified protein fractions were stored at - 80° C until western blotting. Total protein concentration of each sample was determined using a BCA protein assay kit (Pierce, Rockford, IL, USA). For Western blot analysis, 25 µg of protein was loaded into a precast 15-well NuPAGE Novex 12% Bis-Tris gel (Invitrogen, Waltham, MA) for separation by electrophoresis and then transferred to a polyvinylidenedifluoride (PVDF) membrane as indicated in the manufacturer's instructions (GE Healthcare, UK). As primary antibodies, mouse monoclonal anti Aβ antibody (6E10, Covance), mouse monoclonal anti-ß tubulin antibody (cell signaling), rabbit polyclonal antibody to Insulin degrading enzyme (IDE, Abcam ab32216), rabbit anti-amyloid precursor protein C-Terminal antibody (APP-CTF, Sigma A8717) and Neprilysin degrading enzyme (NPE-CD10, Santa Cruz, sc-46656) were incubated at 4° C overnight, followed by secondary horseradish peroxidase-anti-mouse antibody (Jackson Labs, Baltimore) incubation. The blots were visualized with BM chemiluminescence Western blot kit (Roche, Basel).

*ELISA:* Brain protein extracts and plasma were diluted to the fourth into sampling medium provided by the manufacturer and a final of 25µl volume was loaded into a 96 well plates. Aβ38, Aβ40 and Aβ42 measurements were performed according to the manufacturer's instructions (Peptide Panel 1 (6E10) Kit (1 Plate) V-PLEX™K15200E-1 Mesoscale Discovery, Gaithersburg). All levels of Aβ38, Aβ40 and Aβ42 were normalized against total protein amount. For cytokine measurements, total brain homogenates were centrifuged at 25,000g, 30 min, 4°C and supernatants were analyzed using the mouse pro-inflammatory panel 1 V-plex plate (Mesoscale Discovery) according to the manufacturer's instructions and normalized against total protein content.

*Hypothalamic-pituitary-adrenal (HPA) axis:* To evaluate key hormones secreted in the HPA axis, the Millipore MPTMAG-49K MILLIPLEX MAP Mouse Pituitary Magnetic Bead Panel - Endocrine Multiplex Assay was used according to manufacturer's protocol. The analytes available for this 96-well plate multiplex kit are Adrenocorticotropic hormone (ACTH), growth hormone (GH), Prolactin, thyroid stimulating hormone (TSH), luteinizing hormone (LH).

*Statistical analysis (WB & ELISA)*: Data represent the means ± standard errors of the means. Statistical analysis was performed using ANOVA followed by post hoc comparison using Newman-Keuls method.

*Colonization study:* Cecal contents were pooled from amyloid-depositing (12 month-old) CONVR-APPPS1 mice. Cecal contents were homogenized in sterile PBS (2 ml per cecum) and a volume of 0.2 ml was immediately administered by oral gavage to 4 month-old littermate GF-APPPS1 and GF-WT mice, as previously described (*2*). Oral gavage was given at day 1 and day

4. The resulting transplanted mice (termed as COLOAD-APPPS1 and COLOAD-WT) were housed in a conventional environment for 2 months under the same conditions and fed with the same diet as their CONVR-APPPS1 counterparts. Control colonization study was performed using cecal contents pooled from control wild-type (12 month-old) mice and given intraorally to GF-APPPS1 mice (COLOWT-APPPS1). During feces collection, mice were put in clean litter-free individual cages until stool collection. Fresh feces were collected from each mouse at day 1 and 4 and at week 2, 4 and 6 for bacterial 16S rRNA gene sequencing. Mice were analysed 2 months after the beginning of the colonization.

_DNA extraction:_ One frozen faecal pellet/mouse was thawed on ice and DNA was extracted using the QIA amp DNA Stool Mini Kit (Qiagen). The protocol from the manufacturer was followed with an addition of a bead beating step. Sterile glass beads (1 mm) were added together with stool lysis buffer to the samples and cell disruption was performed for 2 ×2 minutes at 25 Hz using a TissueLyser (Qiagen), followed by a heating step at 95℃ for 5 minutes. After lysis, DNA-damaging substances and PCR inhibitors were removed using InhibitEX tablet (provided with the kit) and the DNA was purified on QIAamp Mini spin columns.

_PCR amplification of the V3-4 region of bacterial 16S rRNA genes:_ 16S rRNA genes were amplified by PCR with forward and reverse primers containing Illumina adapter sequences and unique dual indexes used to tag each PCR product (*3*), according to the 16S-protocol provided by Illumina. Primer sequences can be found in Table 1.Briefly, PCR reactions were carried out in 25-μL reactions with 0.2 μM forward and reverse primers, with 12.5 ng template DNA and 12.5 µl of 2× KAPA HiFi HotStart Ready Mix kit (KAPA Biosystems). Thermal cycling consisted of initial denaturation at 95° C for 3 min followed by 25 cycles of denaturation at 95° C for 30 s, annealing at 55° C for 30 s, and extension at 72° C for 30 s, followed by a final step of 72° C for 5 min. The amplicons (464 bp product) were purified with Agencourt AMPureXP Kit (Beckman Coulter). A second PCR was thereafter performed to attach Illumina adapters and unique dual indexes to each sample, followed by a clean-up step with AmPureXP Kit (Beckman Coulter) (**Table s3**). PCR amplicons were visualized using 0.1 % agarose gel electrophoresis. Negative extraction controls did not produce visible bands.

_Amplicons quantitation, pooling, and sequencing:_ Mouse fecal amplicon DNA concentrations were quantified using the Qubit3.0 Fluorometer (Life Technologies, Stockholm, Sweden). Amplicons were combined in equimolar ratios into a single tube with a final concentration of 4pM.As an internal control, 5 % of PhiX was added to the amplicon pool. Paired-end sequencing with a read length of 2x300 bp was carried out on a Miseq Instrument (Illumina) using a Miseq reagent kit v3 (Illumina Inc., San Diego, USA).

_Sequence analysis:_ Sequences were analysed with the free software package Quantitative Insights into Microbial Ecology (QIIME), which allows analysis of high-throughput community sequencing data. Default values were used for each step, except where otherwise specified (*4*). Sequences were removed if lengths were < 200 nt, contained ambiguous bases, primer mismatches or homo polymer runs in excess of six bases. Forward and reverse reads were joined using Fastqjoin. After quality filtering, a total of 6,571,748 sequence reads were generated for the 8 month-old mice (APPPS1 n= 6, WT n= 7), 5,713,081 reads for the 3.5 month-old mice (APPPS1 n= 7, WT n= 8) and 3’688’179 reads for the 1 month-old mice (APPPS1 n= 6, WT n= 8) with an average number of 505,519 reads per sample for the 8 month-old mice, 361,568 reads/sample for the 3.5 month-old and 263,411 reads/sample for the 1 month-old mice. For the colonized mice, a total of 5,550,818 sequence reads was generated after quality filtering with an average of 79,297 reads per sample Similar sequences were binned into operational taxonomic units (OTUs) using UCLUST (*5*), with a minimum pairwise identity of 97 %, generating 1369 OTUs for the CONVR-APPPS1 and WT mice, and 1,837 OTUs for the COLO-mice. The most abundant sequence in each OTU was chosen to represent its OTU. Representative sequences from each OTU were aligned using PyNAST (a python-based implementation of NAST in QIIME) (*3*) and taxonomy was assigned using the Greengenes database (v. 13_8) (*6*) and the RDP classifier (*7*).

*Statistical analysis:* Graphpad Prism 6 software was used to identify significant differences in bacterial relative abundances between transgenic CONVR-APPPS1 and WT mice using multiple t-tests and the Holm-Sidak method of correction for multiple comparisons, at phylum and genus level. To find correlations between brain Aβ42 levels and bacterial genera in the aged and COLO-APPPS1 mice, the operational taxonomic units (OTU) tables were rarefied at 291,712 and 22,609 randomly selected sequences/sample, respectively, for the entire data set after which orthogonal partial least squares (OPLS) scatter plot analysis was performed using SIMCA14 (Umetrics, Umeå, Sweden). Pearson's correlation was calculated for each pairwise combination of brain Aβ42 and bacterial genera using Minitab17 (Minitab Inc, State College, Pennsylvania) and the p-values were then corrected by Benjamini-Hochberg procedure for multiple comparisons (*9,10*) Further, α- and β-diversities were analyzed in QIIME on rarefied data, using 291,712 sequences/sample for 8 month-old mice, 178,620 for 3.5 month-old mice, 136,829 for 1 month-old mice, and 22,609 for COLO-APPPS1 mice. Differences in α-diversity were calculated in QIIME using a non-parametric t-test and FDR correction for multiple comparisons and β-diversity differences were analyzed with the ANOSIM and Adonis non-parametric statistical tests in QIIME. Linear discriminant analysis effect size (LEfSE, www.huttenhower.sph.harvard.edu/galaxy/) was applied on the OTU table according to (*8*).

*Quantification of Thioflavin T-positive amyloid load:* Hemi-brain sections encompassing the cortex and hippocampus, and stained for Thioflavin T (ThT), were quantified for compact amyloid load. To acquire images used for quantification, ThT-stained brain sections were imaged with a 10x objective using a Zeiss Axiovert 200M / ApoTome microscope (Zeiss, Germany) coupled with a Zeiss Axiocam HR camera (Zeiss, Germany). For each of the animals involved in this study, $10 \pm 2$ stained coronal brain secions with a thickness of 25 µm and spaced from each other by 24 slices were available for imaging. Among these, four sections were selected based on the following criteria: one section encompassing the cortex (between position AP 0.74 mm and AP 0.38 mm from Bregma), one section with the striatum (between the position AP - 0.46 mm and AP - 0.70 mm from Bregma), one section showing the dorsal hippocampus (between position AP - 1.82 mm and AP - 2.18 mm from Bregma) and one section encompassing the ventral hippocampus (between position AP - 2.70 mm and AP - 3.08 mm from Bregma). For each selected section, 1300 x 1030 contiguous images (with less than 5 % overlapping) were captured for the entire cortex and hippocampus region (when present). The acquired images were analyzed with the public domain software Image J. Analyzed areas were adjusted manually so that each measurement was accurately measuring only the region of interest and adjusted to exclude brains regions other than the cortex and hippocampus. For ThT-load analysis, the images were inverted and a threshold of approximately 180 was applied (with a 5% variance to best adjust tissue staining variations among sections). ThT-stained brain sections were analyzed with parameters of larger than 5x5 pixels and with a circularity of 0.15. Statistics were processed using Statistica (StatSoft, Tulsa, OK) and included analysis of variance and posthoc tests.

*Quantification of Iba-1-positive immunoreactivity:* Neocortical microglia immunoreactivy was quantified in hemi-brain sections immunostained for activated microglia (with the Iba-1 antibody). Acquisition of images from Iba-1-stained sections was similar to that described for ThT quantification. Subsequently, a threshold of 110 with a variance of 10% was applied to each image using Image J. This threshold was selected in order to best display the microglia while minimizing the background noise. After visual inspection, ROI were selected such that staining artefacts on the section was manually removed by cropping. The microglia were quantified using the analyze particles tool embedded in ImageJ. In order to consider both the single microglia and the microglial loads, the pixel size was not specified and the circularity parameter was set to 0.0-1.0. Data were then processed using Statistica.

*References (methods):*

**Acknowledgments:** The financial contributions of the following foundations are well acknowledged: Gebert Rüf foundation (grant GRS-002/13) (T.B.) and the Direktör Albert Påhlsson Foundation and the Mats Paulsson Foundation for research, innovation and societal development (N.M. and F.F).

**Author contributions**: T.H and V.C. performed biochemical experiments. N.D performed immuno-histological experiments. N.M and F.F assessed gut microbiota profile and performed bacterial 16S rRNA sequencing. J.J.N. performed cytokine assessments. K.D.M generated and maintained germ free mouse model. T.L, T.H., F.F, M.J. and T.B wrote the manuscript.

**Author information**:
Reprints and permissions information is available at www.nature.com/reprints.
All authors declare no competing financial interest.
Correspondence and requests for materials should be addressed to Professor T. Lasser (theo.lasser@epfl.ch), or Dr T. Bolmont (tristan.bolmont@epfl.ch).

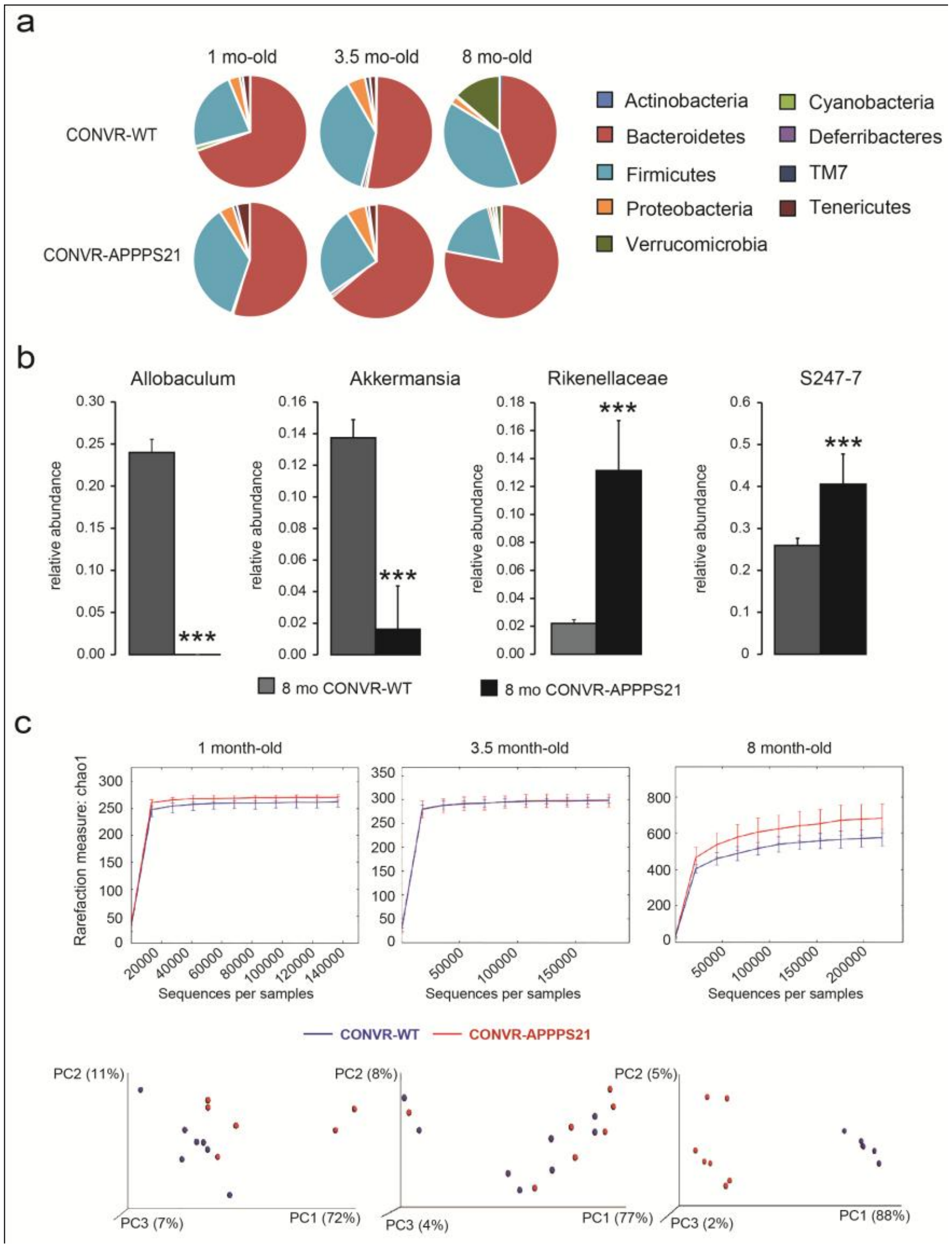

**Fig. 1.** Comparison of the gut microbiota between transgenic conventionally raised (CONVR-) APPPS1 and wild type (WT) mice analyzed with multiple t-tests together with the Holm-Sidak method to correct for multiple comparisons. **(a)** Mean sequence relative abundance of gut microbial taxa at phylum level in CONVR-APPPS1 and WT mice, aged 1 month (n= 6 and n= 8 respectively), 3.5 months (n= 7 and n= 8) and 8 months (n= 7 and n= 6). Bacteroides, Firmicutes, Verrucomicrobia and Tenericutes were significantly different between 8 month-old APPPS1 and WT mice ($p < 0.001$), as well as Proteobacteria ($P$=0.0016) and Actinobacteria ($P$=0.0053). No significant differences were found in the younger mice. **(b)** At genus level, 4 microbial taxa with relative abundance > 5 % were significantly different between 8 month-old CONVR-APPPS1 and WT mice (***$p < 0.001$), while younger mice did not show any significant differences. **(c)** Rarefaction curves (α-diversity vs. sequencing effort) and weighted Unifrac PCoA plot to compare phylogenetic distance matrices of CONVR-APPPS1 and WT mice. Species richness was significantly increased in the 8 month-old APPPS1 mice ($P$=0.033, observed species test in QIIME with FDR correction for multiple comparisons). Younger mice did not show significant differences in α-diversity. Clustering of CONVR-APPPS1 mice in the PCoA was significantly separated from the WT mice at 8 months of age ($p < 0.001$ for both the ANOSIM and the Adonis non-parametric test in QIIME).

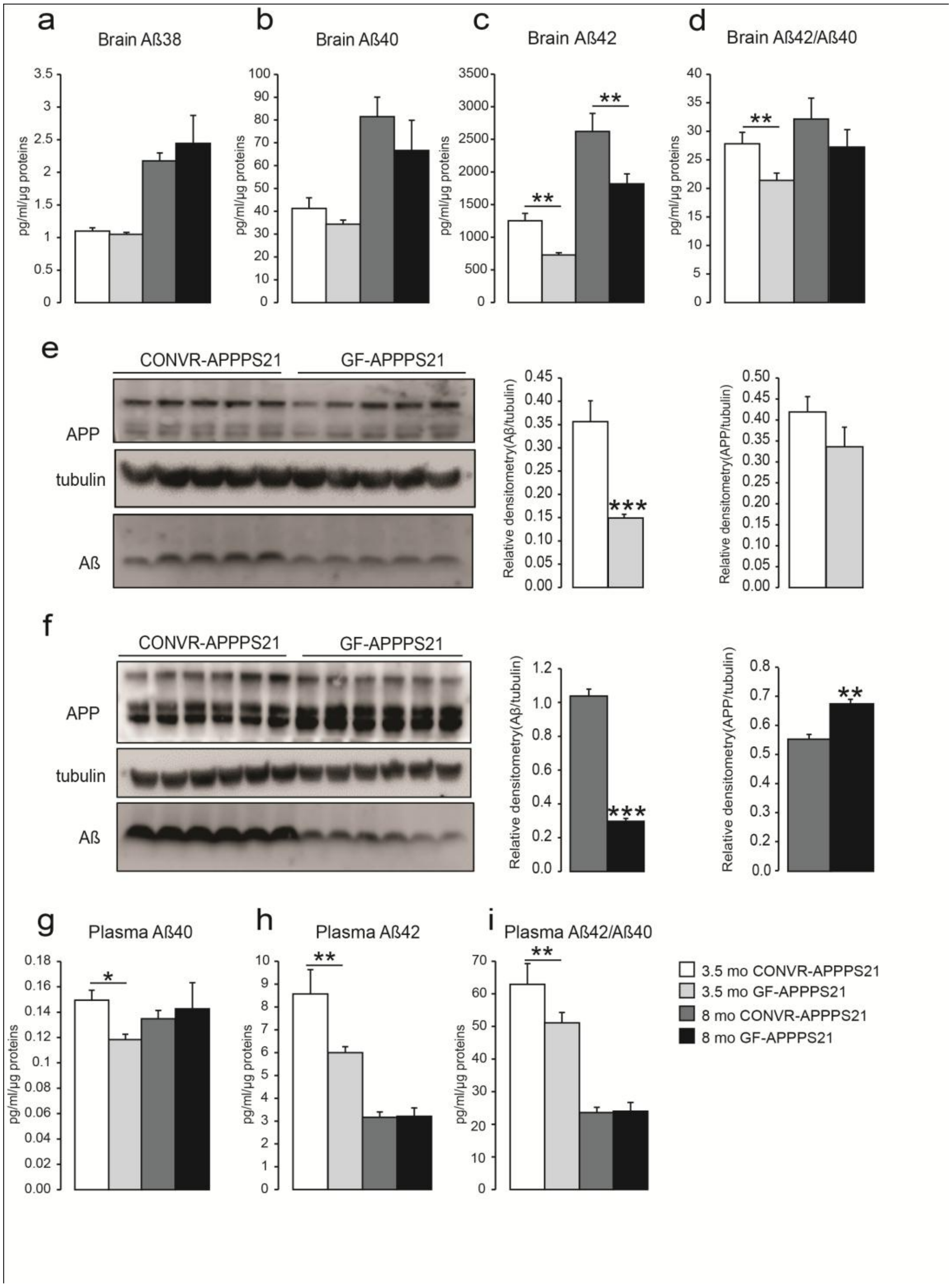

**Fig. 2.** Reduction of cerebral and plasmatic soluble Aβ levels in GF-APPPS1 transgenic mice. Levels of cerebral soluble Aβ38 **(a)**, Aβ40 **(b)**, Aβ42 **(c)** and ratio Aβ42/ Aβ40 **(d)** by ELISA in 3.5 (n= 7) and 8 (n= 7) month-old conventionally-raised (CONVR-APPPS1) and germ free (GF-APPPS1) APPPS1 mice. Levels of Aβ assessed by western blot in 3.5 month-old (n= 5) **(e)** and in 8 month-old mice (n= 6) **(f)**. Plasmatic levels of soluble Aβ40 **(g)**, Aβ42 **(h)** and ratio Aβ42/Aβ40 **(i)** by ELISA. Data represent mean ± SEM. Statistical differences between germ free and conventionally-raised mice: *: $p < 0.05$, **: $p < 0.01$, ***: $p < 0.001$.

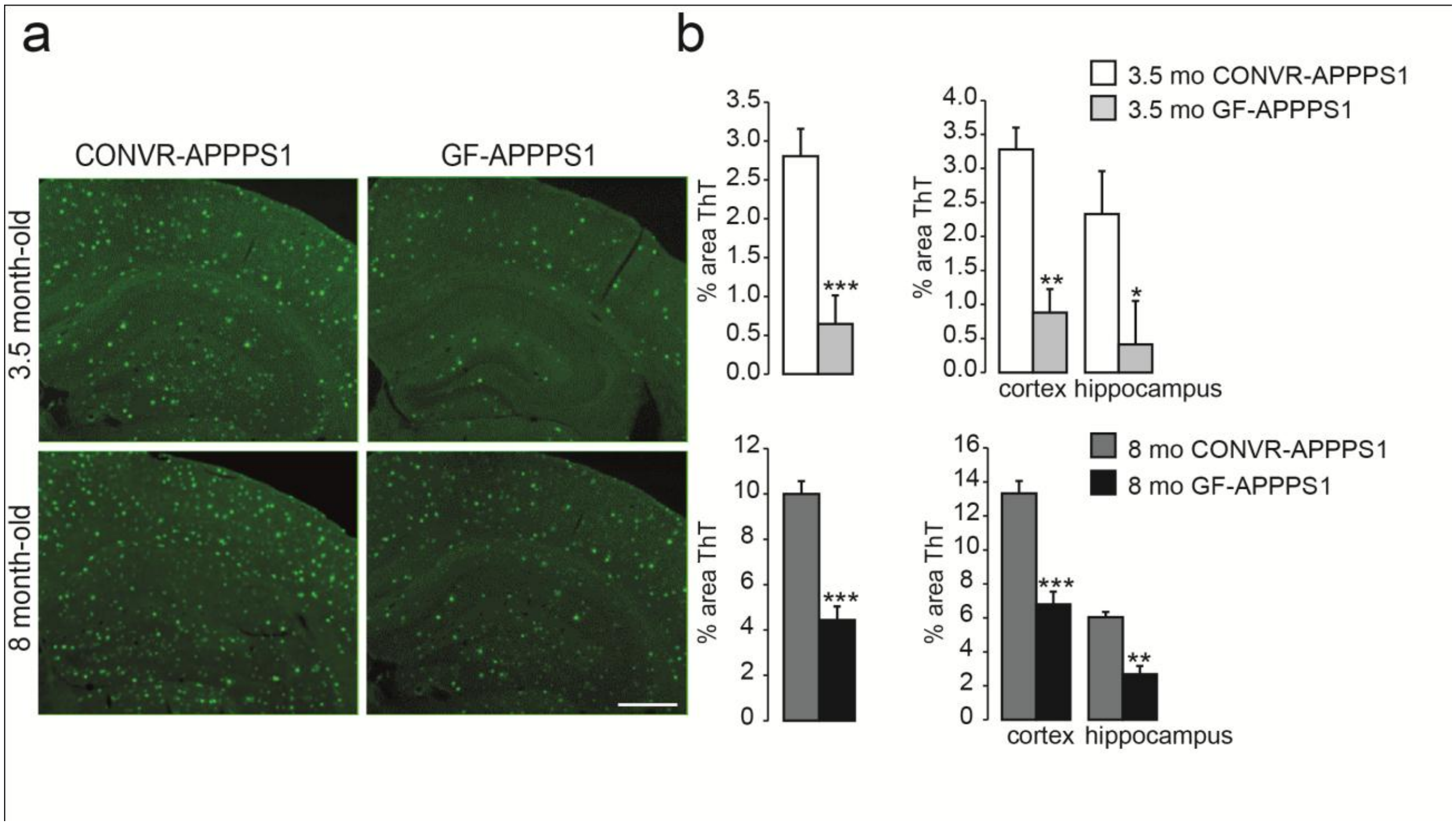

**Fig. 3.** Decreased cerebral Aβ amyloidosis in GF-APPPS1 transgenic mice. **(a)** Representative Thioflavin T-stained brain sections encompassing the cortex and hippocampus of a 3.5 month-old CONVR-APPPS1 mouse (top left) as compared to an age-matched GF-APPPS1 mouse (top right). Compact Aβ amyloid pathology in a 8 month-old CONVR-APPPS1 mouse (bottom left) as compared to an age-matched GF-APPPS1 mouse (bottom right) (scale bar: 150 um, all panels have the same magnification). **(b)** Quantification of amyloid load on Thioflavin-stained sections demonstrates a significant reduction of cerebral Aβ amyloid plaques in young and aged GF-APPPS1 mice (n= 7 for both groups) in both cortex and hippocampus, as compared to CONVR-APPPS1 mice (n= 7 for both groups). Statistical differences between GF- and CONVR-APPPS1 mice: *: $p < 0.05$, **: $p < 0.01$, ***: $p < 0.001$. Shown are mean ± SEM.

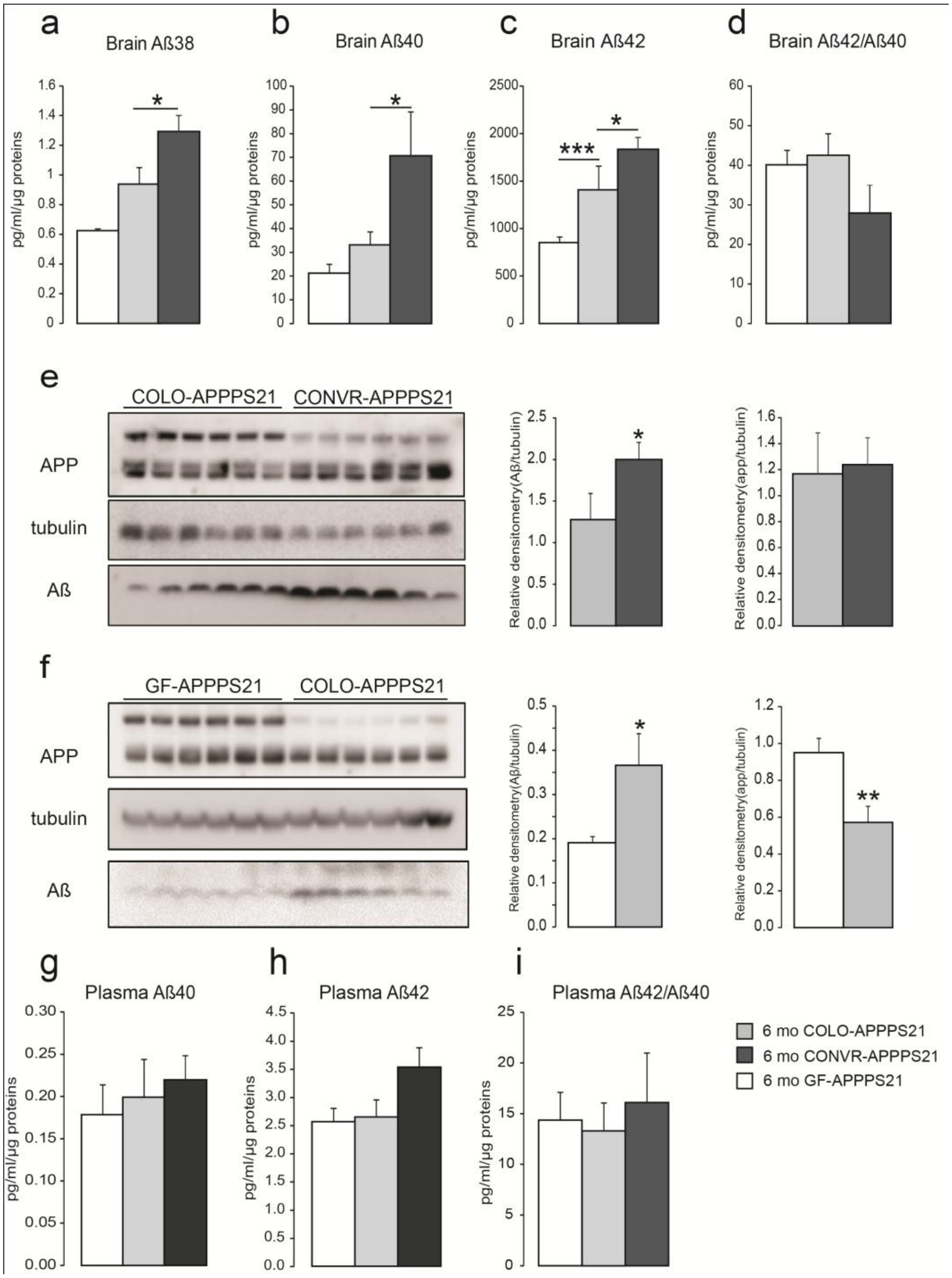

**Fig. 4.** Colonization of GF-APPPS1 mice with the microbiota from aged CONVR-APPPS1 (COLOAD-APPPS1). Levels of cerebral soluble Aβ38 **(a)**, Aβ40 **(b)**, Aβ42 **(c)** and ratio Aβ42/Aβ40 **(d)** by ELISA in 6 month-old GF-APPPS1, COLOAD-APPPS1 and CONVR-APPPS1 mice. Levels of Aβ assessed by western blot in 6 month-old COLOAD-APPPS1 and CONVR-APPPS1 mice (n= 6) **(e)** and in 6 month-old GF-APPPS1 and COLOAD-APPPS1 animals (n= 6) **(f)**. Plasmatic levels of soluble Aβ40 **(g)**, Aβ42 **(h)** and ratio Aβ42/Aβ40 **(i)** by ELISA. Data represent mean ± SEM. Statistical differences between germ free and conventionally-raised mice: *: $p < 0.05$, **: $p < 0.01$, ***: $p < 0.001$.

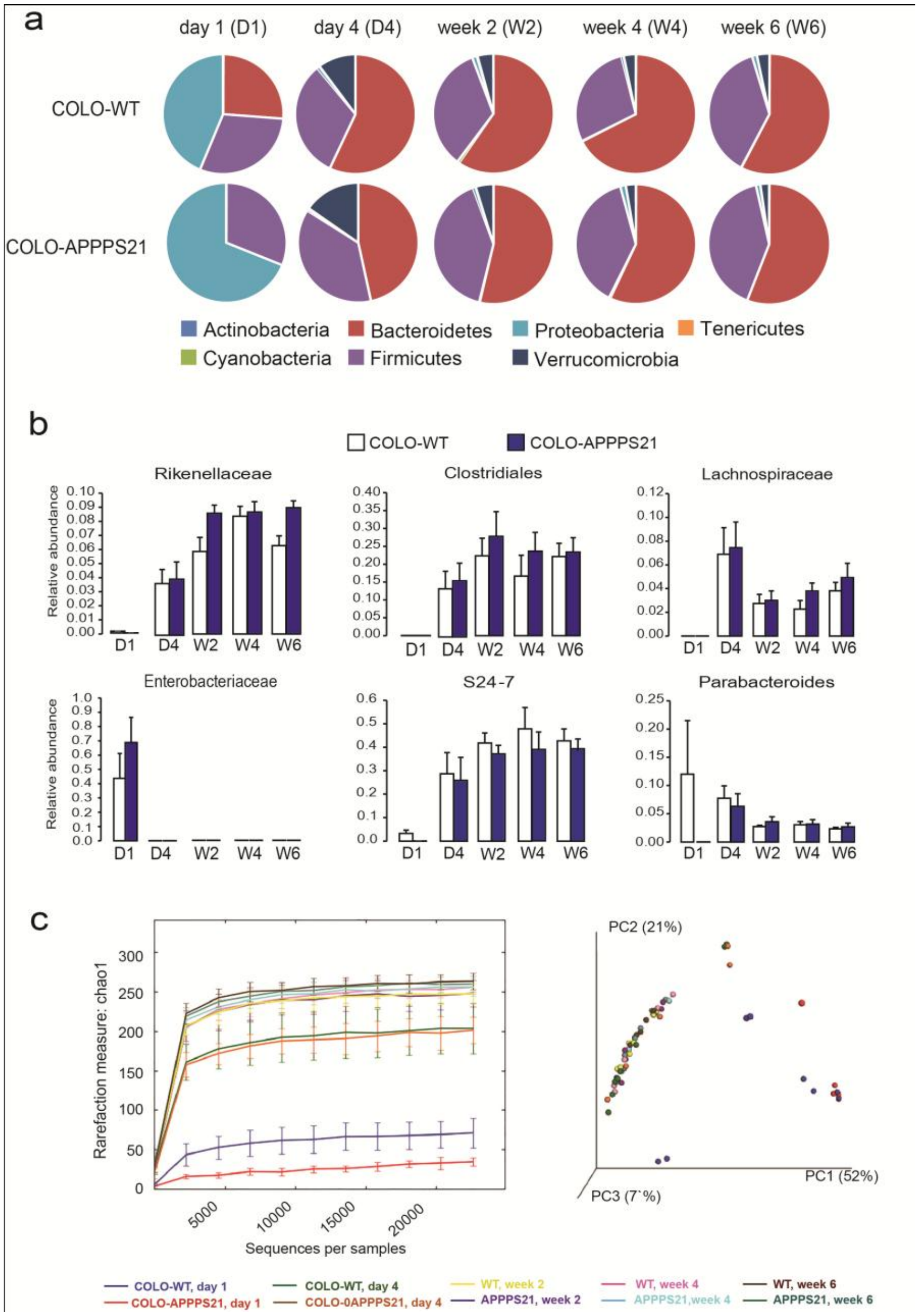

**Fig. 5.** Comparison of the gut microbiota between animals in colonization experiments. Mean sequence relative abundance of gut microbial taxa at phylum level in COLO-APPPS1 and COLO-WT mice (n=6 and n=8, respectively) at day 1 and 4 after colonization and at week 2, 4 and 6 (a). Relative abundance of high-abundant (> 5%) bacterial taxa at genus level in COLO-WT mice and in COLO-APPPS1 mice (b). Rarefaction curves (α-diversity vs. sequencing effort) (left panel) and weighted Unifrac PCoA plot to compare phylogenetic distance matrices (right panel) of COLO-APPPS1 and WT mice at different time points. The principal components (PC) describes the % variation in phylogenetic diversity that is explained by the PC. Species richness did not differ between COLO-WT mice as compared to COLO-APPPS1 mice at any of the time points (observed species test in QIIME with FDR correction for multiple comparisons).Clustering of COLO-APPPS1 mice in the PCoA was significantly separated from the WT mice at day 1 ($p< 0.001$ for both the ANOSIM and the Adonis non-parametric test in QIIME) (c).

## Supplementary Figures

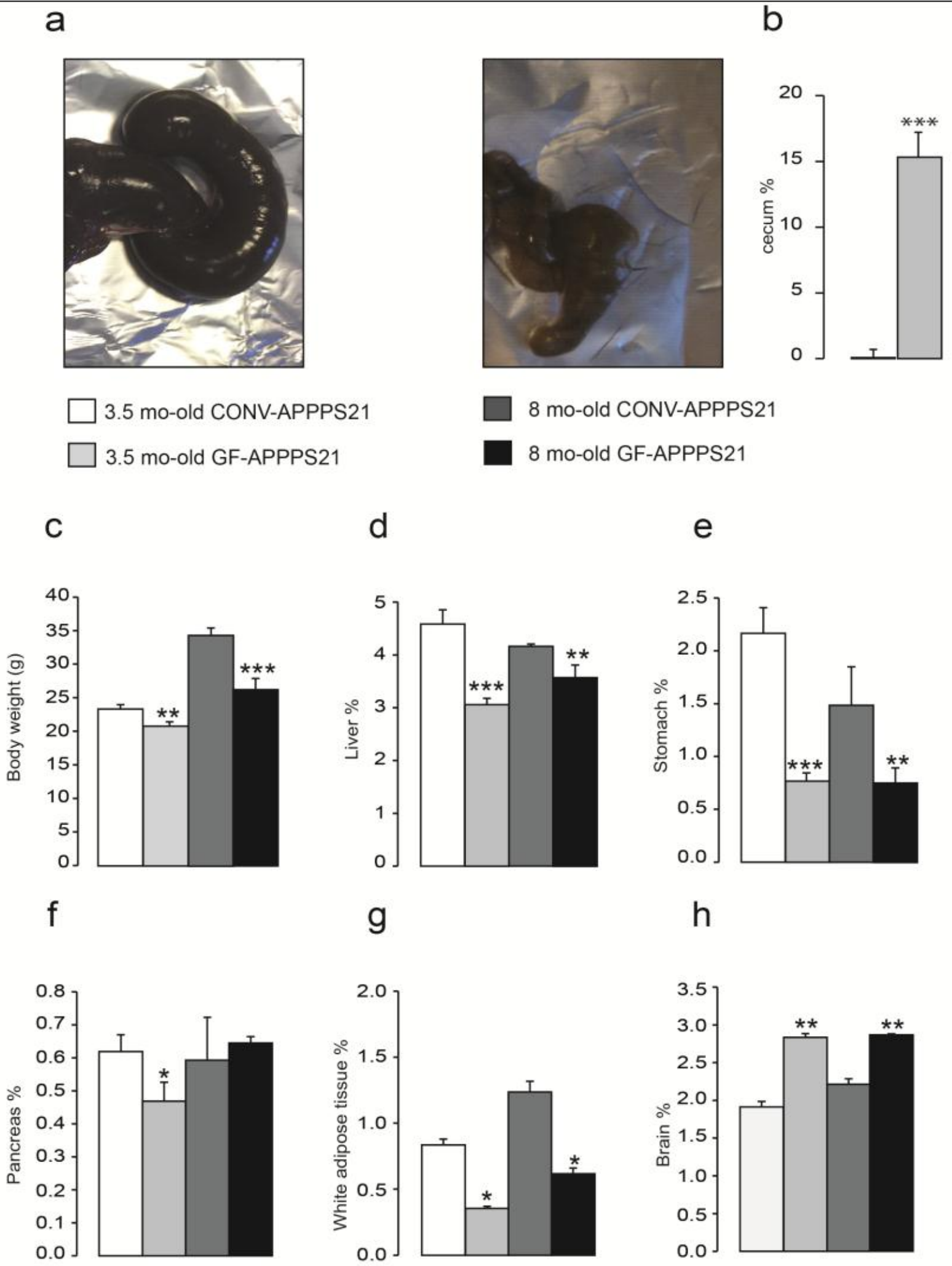

**Fig. s1**. Metabolic parameters of young and aged APPPS1 mice. Cecal photograph of representative 3.5 month-old GF- (left panel) and age-matched (right panel) CONVR-APPPS1 mice **(a)**. Cecal weight of 3.5 month-old GF- and 3.5 month-old CONVR-APPPS1 mice **(b)**. Body **(c)**, liver **(d)**, stomach **(e)**, pancreas (**f)**,white adipose tissue (WAT) **(g)** and brain weights **(h)** of young (3.5 month-old) and aged (8 month-old) CONVR- APPPS1 and GF-APPPS1 mice. Data represent mean ± SEM. Statistical differences between GF- and CONVR-APPPS1 mice: *: p< 0.05, **: p< 0.01, ***: p< 0.001.

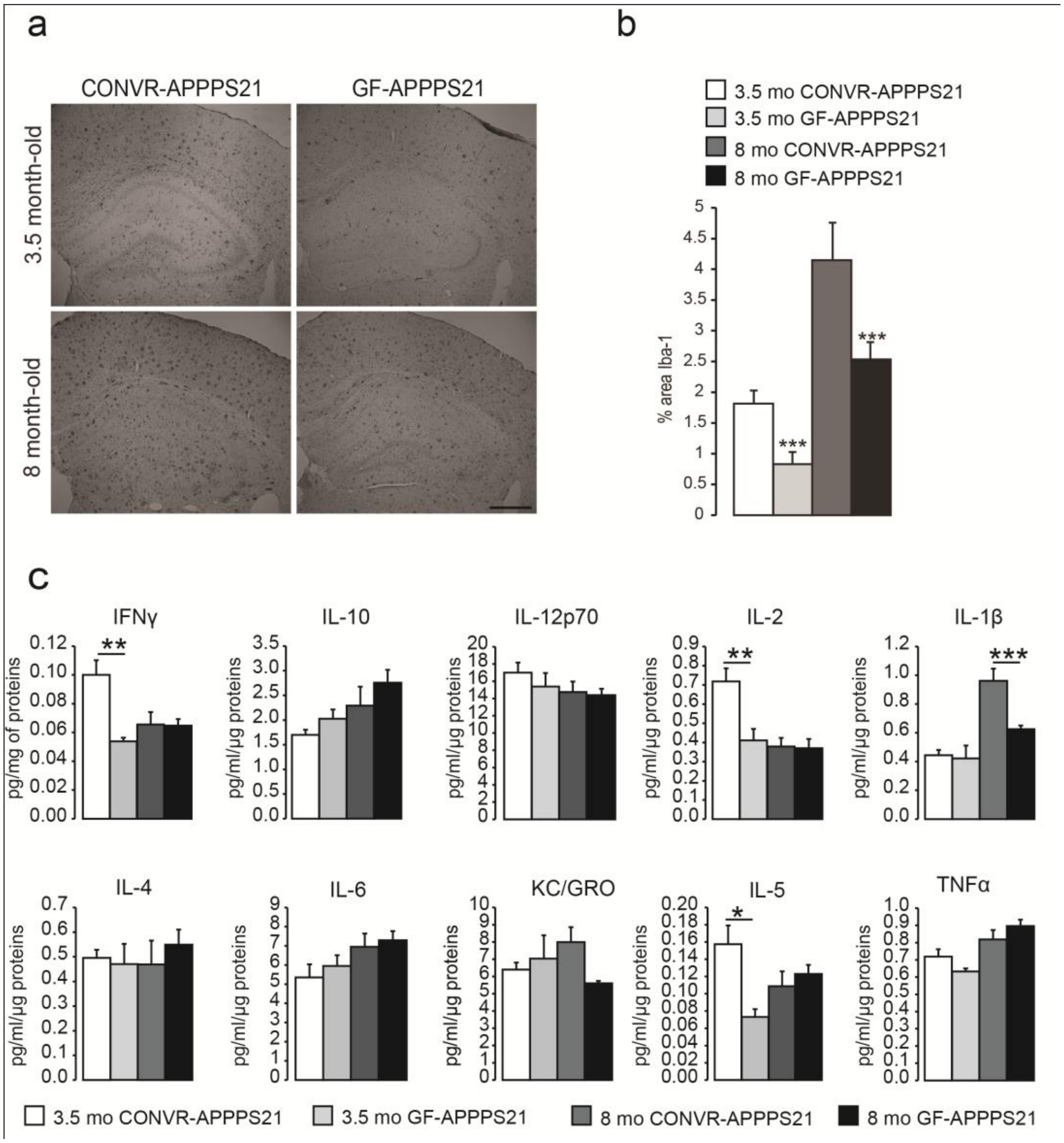

**Fig. s2**. Decreased neuroinflammation in GF-APPPS1 mice. Representative sections immunostained with Iba-1 in young and aged CONVR-APPPS1 mice (left panels) and GF-APPPS1 mice (right panels) (scale bar: 150 um, all panels have the same magnification) **(a)**. Quantification of Iba-1 immunoreativity demonstrates a significant reduction of activated neocortical microglia in GF-APPPS1 mice as compared to CONVR-APPPS1 animals. Statistical differences between GF- and CONVR-APPPS1 mice: *: $p < 0.05$, **: $p < 0.01$, ***: $p < 0.001$. Shown are mean ± SEM **(b)**. Cytokine profiles in GF-APPPS1 and CONVR-APPPS1 mice. 3.5 month-old GF-APPPS1 mice show significantly reduced levels of interferon (IFN)-γ, interleukin (IL)-2 and IL-5 compared to CONR-APPPS1 animals. Further, while there is a significant increase of IL-1β with age in CONVR-APPPS1 mice, IL-1β levels are significantly lower in 8 month-old GF animals. Other cytokines, namely IL-10, IL-12p70, IL-4, IL-6 (**g**), KC/GRO and TNF-α show no significant alterations in GF compared to CONVR-APPPS1 mice. * $p < 0.05$, ** $p < 0.01$, *** $p = 0.001$ **(c)**.

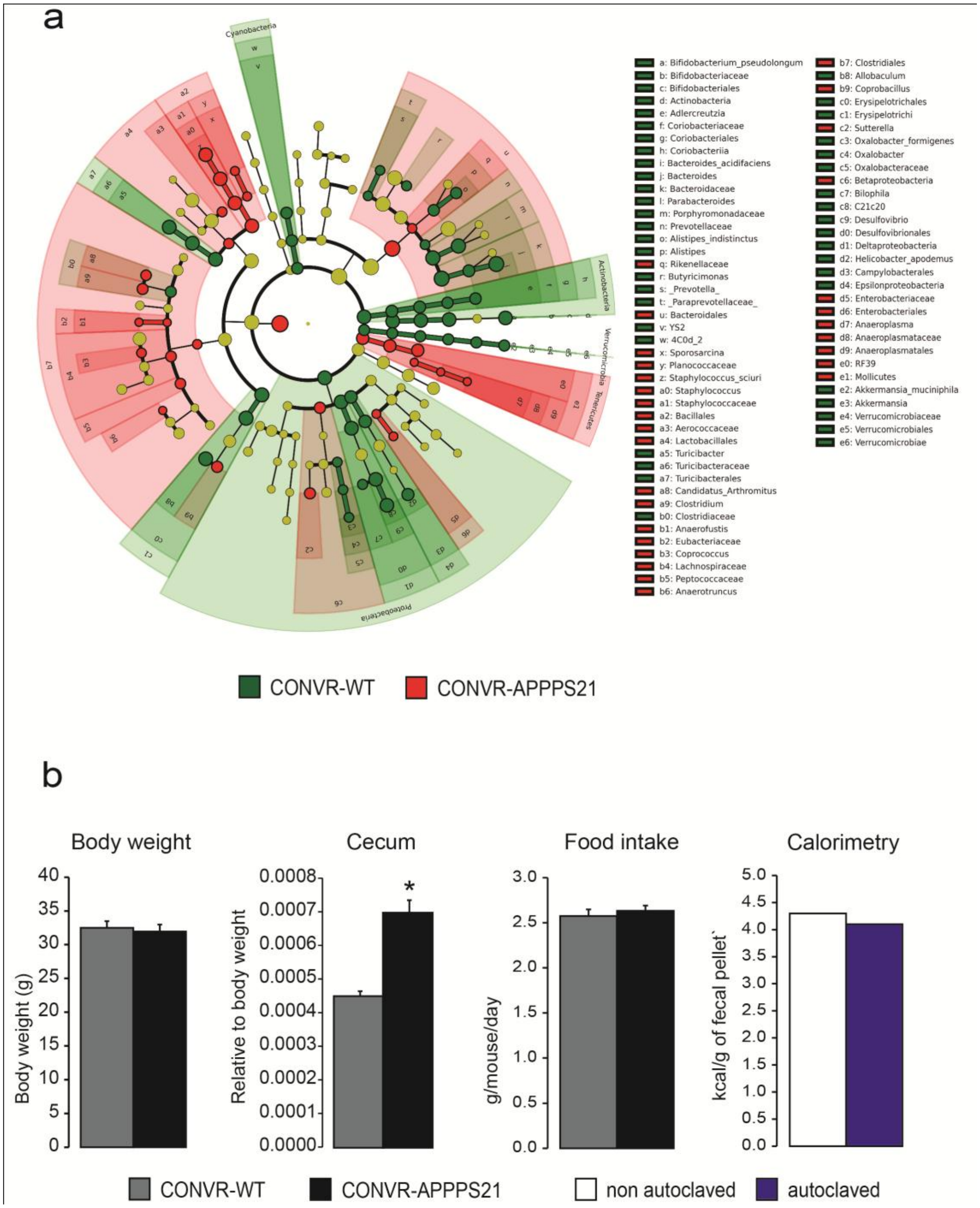

**Fig. s3.** Comparison of the gut microbiota between 8 month-old transgenic (CONVR-APPPS1) and wild type (CONVR-WT) mice. Cladogram generated with LDA Effect Size (LEfSE) analysis ($\alpha$=0.01) displays bacterial taxa with LDA score higher than 2, illustrating gut microbial taxa present in higher abundance in 8 month-old CONVR-APPPS1(red) and CONVR-WT (green)mice **(a)**. Size of circles is proportionate to each taxon's mean relative abundance. Body weight and cecum weight in 8 month-old CONVR-APPPS1 and CONVR-WT mice. Data represent mean ± SEM. Statistical differences between wild type and APPPS1 transgenic mice: * $p < 0.05$ **(b)**.

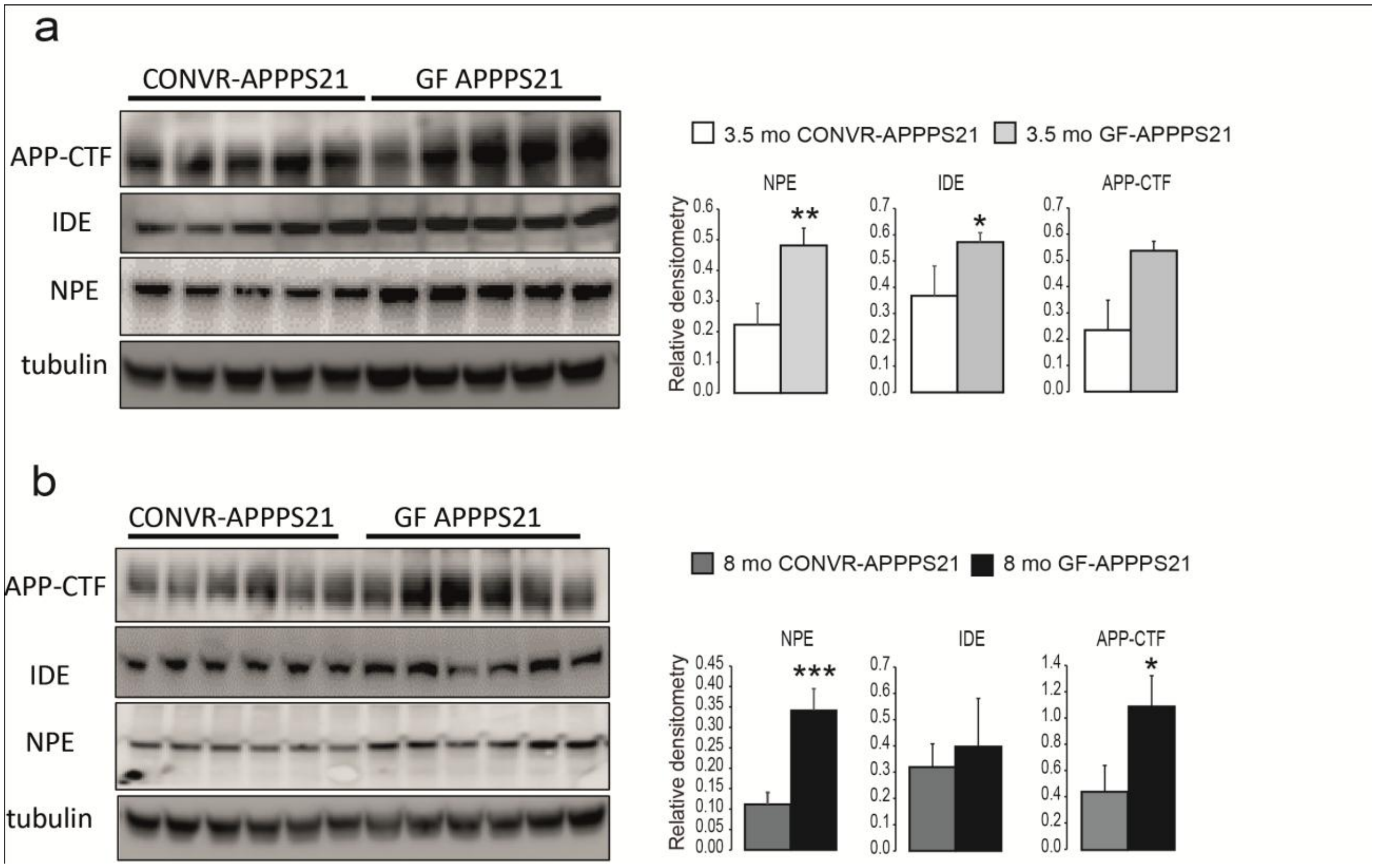

**Fig. s4**. Levels of Neprilysin degrading enzyme (NPE), Insulin degrading enzyme (IDE) and APP-CTF in the brain of 3.5 **(a)** and 8 month-old **(b)** GF-APPPS1 and CONR-APPPS1 mice. Levels of NPE were increased in young and aged GF-APPPS1 animals compared to age-matched CONVR-APPPS1 mice. Levels of IDE were increased in young GF-APPPS1 animals compared to age-matched CONVR-APPPS1 mice, but similar between aged animals. Levels of APP-CTF did not differ significantly between young animals but were increased in aged GF-APPPS1 animals compared to age-matched CONV-APPPS1 mice.

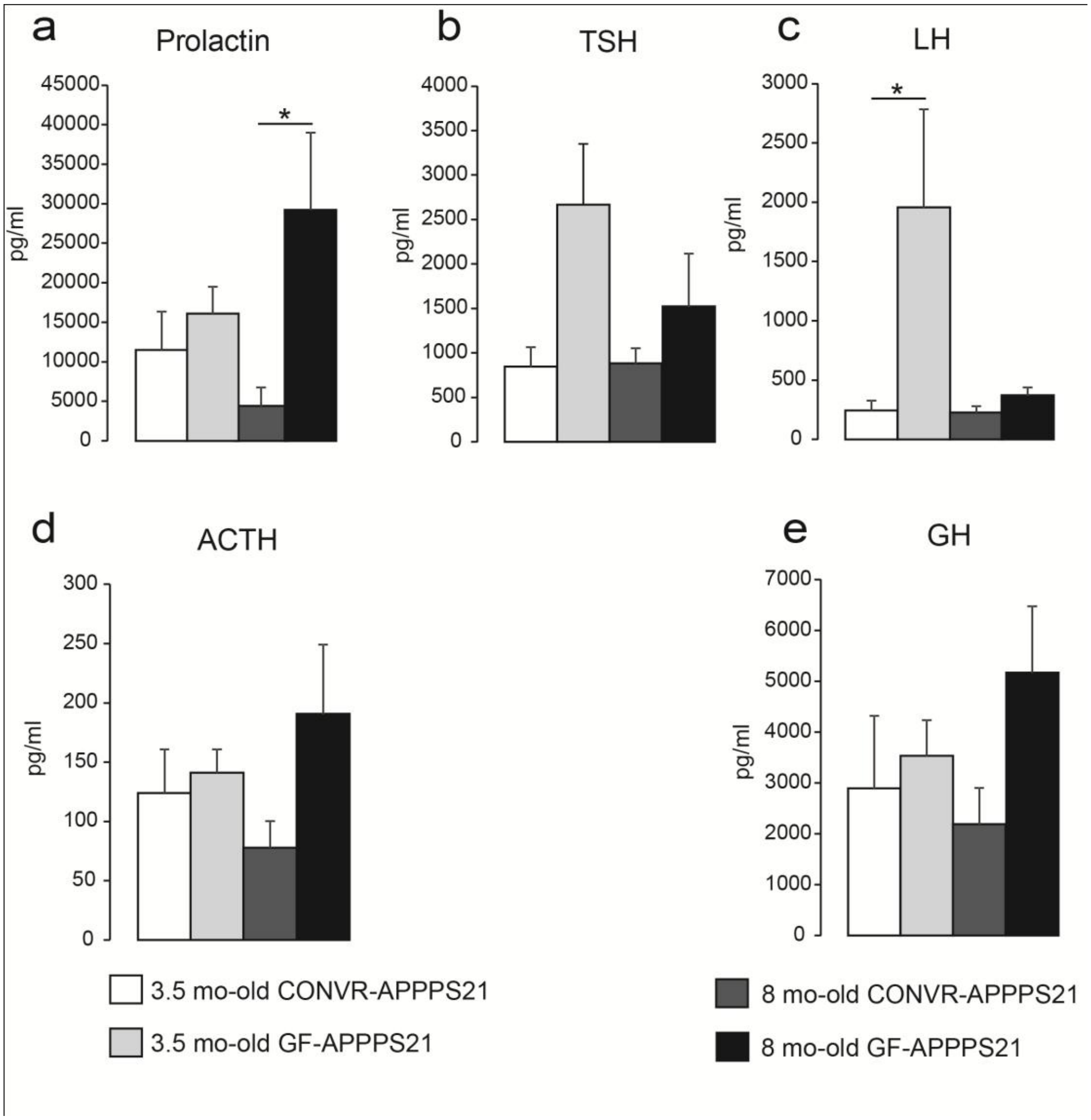

**Fig. s5**. Neuroendocrine levels in the hypothalamic-pituitary-adrenal (HPA) axis. Levels of Prolactin **(a)**, thyroid stimulating hormone (TSH) **(b)**, luteinizing hormone (LH) **(c)** Adrenocorticotropic hormone (ACTH) **(d)**, growth hormone (GH) **(e)** in 3.5 and 8 month-old GF-APPPS1 and CONVR-APPPS1 mice.

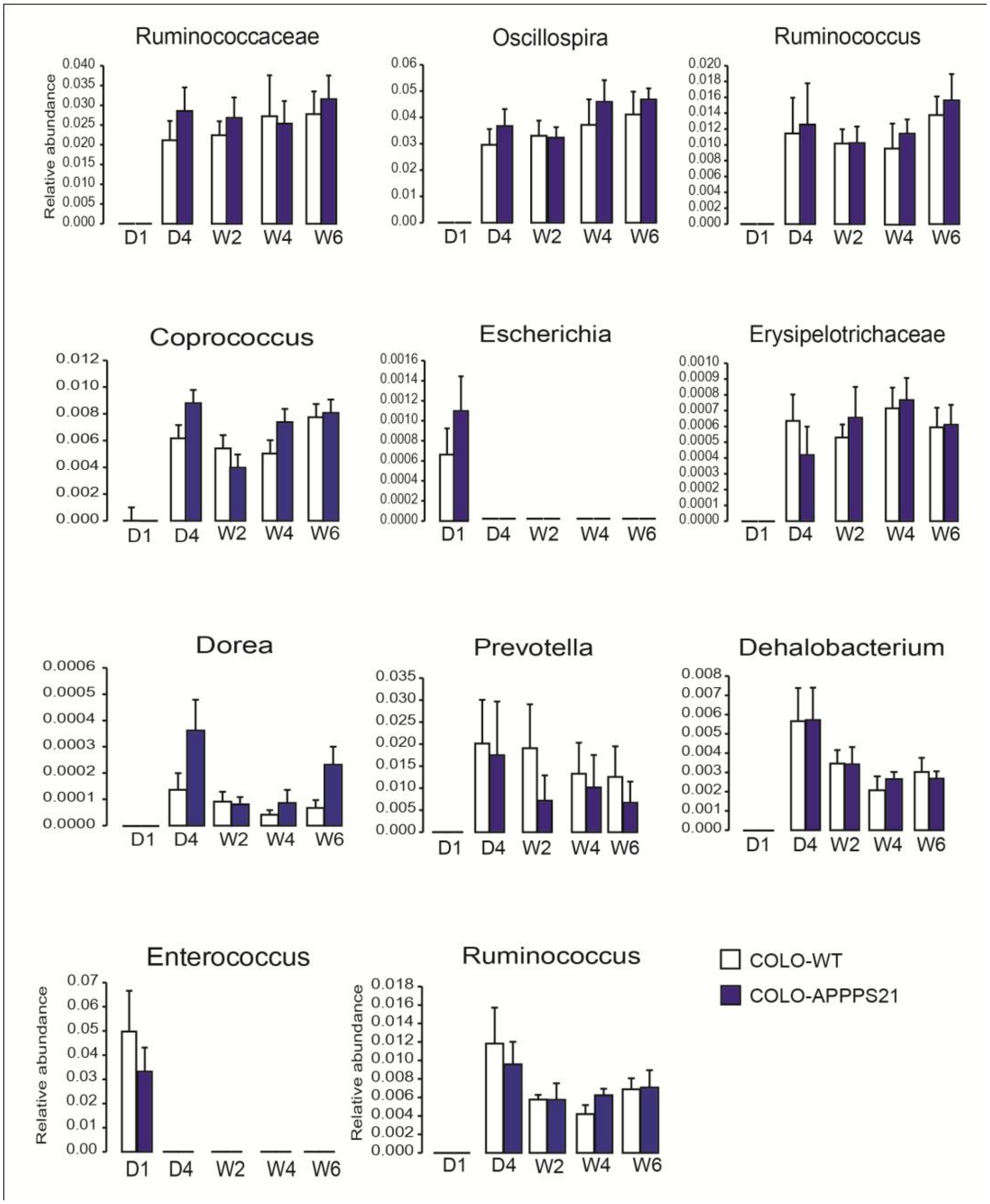

**Fig. s6.** Mean sequence relative abundance of low-abundant (<5%) bacterial taxa at genus level in COLO-APPPS1 and COLO-WT mice. No significant differences between COLO-APPPS1 and COLO-WT mice was observed at any of the time points. Comparing abundance at day 1 (D1) with the other time points within the COLO-APPPS1 and COLO-WT mice, respectively, revealed significant changes in bacterial abundance over time. * $p < 0.05$, ** $p < 0.01$, *** $p < 0.001$.

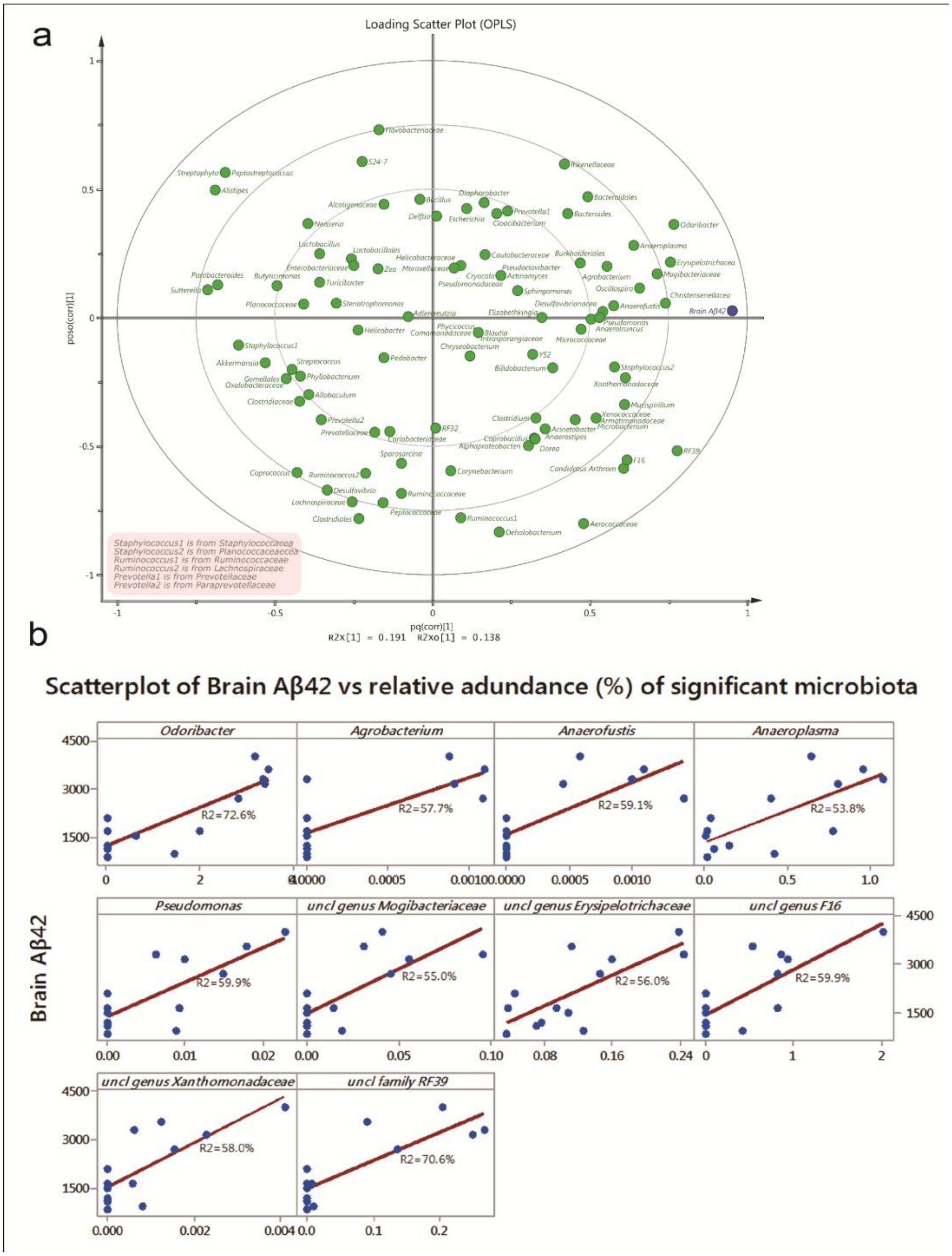

**Fig. s7.** Correlation between the gut microbiota and cerebral soluble Aβ42 in 8 month-old CONVR-APPPS1 and COLO-APPPS1 mice. Orthogonal partial least squares (OPLS) scatter plot giving an overview of the correlations. Variables situated closed to each other were positively correlated and variables situated opposite each other were negatively correlated (a). **(b)** . Ten bacterial genera were positively correlated with Aß42 levels ($p < 0.05$, Pearson's correlation with adjusted p-values for multiple comparisons using Benjamini-Hochberg procedure). Scatter plot of the ten significant correlations with determination coefficient ($R^2$) indicating how well the data fits the model.

## Supplementary Tables

| | | Aerobic culture ceacal content | Anaerobic culture ceacal content | Gram stain ceacal content | Cytox DNA stain ceacal content |
|---|---|---|---|---|---|
| 8.1.2014 | Isolator BE.09 Sentinel #1 | neg | neg | neg | neg |
| 8.1.2014 | Isolator BE.09 Sentinel #2 | neg | neg | neg | neg |
| 7.2.2014 | Isolator BE.09 Sentinel #1 | neg | neg | neg | neg |
| 7.2.2014 | Isolator BE.09 Sentinel #2 | neg | neg | neg | neg |
| 8.4.2014 | Isolator BE.09 Sentinel #1 | neg | neg | neg | neg |
| 8.4.2014 | Isolator BE.09 Sentinel #2 | neg | neg | neg | neg |

**Table S1.** Aerobic, anaerobic culture, sytox-green and gram staining of germ free cecal content. "neg" refers to negative detection at mentioned time points.

| Organism | Sample | Test Method | 20.03.2014 | 08.05.2014 |
|---|---|---|---|---|
| ***Viruses*** | | | | |
| Ectromelia virus | Serum | IFA | 0/2 | 0/2 |
| Lymphocyticchoriomeningitis (LCMV) | Serum | ELISA | 0/2 | 0/2 |
| Mouse adenovirus, FL | Serum | ELISA | 0/2 | 0/2 |
| Mouse adenovirus, K87 | Serum | ELISA | 0/2 | 0/2 |
| Mouse cytomegalovirus (MCMV) | Serum | IFA | 0/2 | 0/2 |
| Mouse Rotavirus (EDIM) | Serum | IFA | 0/2 | 0/2 |
| Mouse hepatitis virus (MHV) | Serum | nPCR | 0/2 | 0/2 |
| MurineNorovirus (MNV) | Serum | IFA | 0/2 | 0/2 |
| Parvoviruses: Mouse parvovirus (MPV) | Serum | IFA | 0/2 | 0/2 |
| Parvoviruses: minute virus of mice (PVM) | Serum | IFA | 0/2 | 0/2 |
| Reovirus type 3 (REO 3) | Serum | IFA | 0/2 | 0/2 |
| Sendai virus | Serum | IFA | 0/2 | 0/2 |
| Theiler'smurineencephalomyelitis (TMEV) | Serum | IFA | 0/2 | 0/2 |
| Theiler (GDVII) | Serum | IFA | 0/2 | 0/2 |
| ***Mycoplasma*** | | | | |
| Mycoplasma pulmonis | Serum | IFA | 0/2 | 0/2 |
| ***Bacteria*** | | | | |
| Citrobacter rodentium | RLGImt* | Culture | 0/2 | |
| Clostridium piliforme | Serum | IFA | 0/2 | 0/2 |
| Corynebacterium kutscheri | RLGImt* | Culture | | 0/2 |
| Helicobacter sp. | Caecum | nPCR | 0/2 | |
| Pasteureillaceae | RLGImt* | Culture | 0/2 | |
| Salmonella spp. | RLGImt* | Culture | 0/2 | |
| Streptoba caillusmoniliformis | RLGImt* | Culture | 0/2 | |
| Streptococcus pneumoniae | RLGImt* | Culture | 0/2 | |
| beta-haemolytische Streptococcen | RLGImt* | Culture | 0/2 | |
| Bordetella bronchiseptica | RLGImt* | Culture | 0/2 | |
| Klebsiella oxytoca | RLGImt* | Culture | 0/2 | |
| Staphylococcus aureus | RLGImt* | Culture | 0/2 | |
| Pseudomonas sp. | RLGImt* | Culture | 0/2 | |
| Pseudomonas aeruginosa | RLGImt* | Culture | 0/2 | |
| Yersiniapseudo tuberculosis | RLGImt* | Culture | 0/2 | |
| Pasteurellapneumotropica | Serum | IFA | 0/2 | 0/2 |
| ***Parasites*** | | | | |
| Ecotoparasites | Skin/Pelt | Microscopy | 0/2 | |
| Endoparasites-Protozoa | Native test | Microscopy | 0/2 | |
| Helminths | Caecum | Microscopy | 0/2 | |
| Toxoplasmagondii | Serum | IFA | 0/2 | |

*RLGImt* :* *Respiratory Tract/Lung,GI Tract, external mucosa, throat*
*nPCR:* *PCR non GLP*

**Table S2.** Serological testing for common viruses and pathogens.

| | |
|---|---|
| 16S Amplicon PCR Forward Primer | 5'CCTACGGGNGGCWGCAG |
| 16S Amplicon PCR Reverse Primer | 5'GACTACHVGGGTATCTAATCC |
| Illumina Forward overhang: | 5’ TCGTCGGCAGCGTCAGATGTGTATAAGAGACAG-16S-V3-4-specific sequence |
| Illumina Reverse overhang: | 5’ GTCTCGTGGGCTCGGAGATGTGTATAAGAGACAG-16S-V3-4-specific sequence |

**Table S3.** Primer sequences for amplification and sequencing of 16S rRNA genes.